\def\plb#1{Phys.~Lett.~{\bf B#1}}
\def\npb#1{Nucl.~Phys.~{\bf B#1}}
\def\prl#1{Phys.~Rev.~Lett.~{\bf #1}}
\def\prd#1{Phys.~Rev.~{\bf D#1}}

\def\a{\alpha}

\def\l{\left}
\def\r{\right}
\def\ord#1{{\cal O}\l(#1\r)}
\def\la{\langle}
\def\ra{\rangle}

\def\im{\mbox{Im}\,}

\def\tab#1{Table~\ref{#1}}

\def\fig#1{Fig.~\ref{#1}}

\def\sec#1{Sec. \ref{#1}}
\def\eq#1{Eq.~(\ref{#1})}

\def\re#1{Ref.~\cite{#1}}

\documentstyle[twoside,fleqn,espcrc2,epsf]{article}

\newcommand{\AmS}{{\protect\the\textfont2
  A\kern-.1667em\lower.5ex\hbox{M}\kern-.125emS}}

\newdimen\unit
\def\point#1 #2 #3{\vbox to0pt{\kern-#2\unit
  \hbox{\kern#1\unit#3}\vss}
 \nointerlineskip}

\newcommand{\be}{\begin{equation}}
\newcommand{\ee}{\end{equation}}
\newcommand{\bea}{\begin{eqnarray}}
\newcommand{\eea}{\end{eqnarray}}

\newcommand{\gev}{\mbox{ GeV}}

\newcommand{\plus}{\makebox[12pt][c]{$+$}}
\newcommand{\minus}{\makebox[12pt][c]{$-$}}

\newcommand{\er}[2]{\raisebox{0.08em}{\scriptsize {$\begin{array}{@{}l@{}}
                          \plus\makebox[0.15em][r]{#1} \\[-0.12em] 
                          \minus\makebox[0.15em][r]{#2} 
                        \end{array}$}}}

\title{Semileptonic $B\to\rho$ and $B\to\pi$ Decays: Lattice and Dispersive
Constraints}

\author{Laurent Lellouch\address{Centre de Physique Th\'eorique,
CNRS-Lumniy,Case 907, F-13288 Marseille Cedex 9, France}
\thanks{CPT is UPR 7061. The $B\to\rho\ell\bar\nu$ 
project was supported by SERC grants GR/G32779
and GR/H49191 and PPARC grant GR/J21347. I am grateful to my colleagues
of the UKQCD Collaboration for many fruitful discussions. }}

\begin{document}


\begin{titlepage}

\begin{center}

\renewcommand{\thefootnote}{\fnsymbol{footnote}}

{\Large\bf Centre de Physique Th\'eorique\footnote{
Unit\'e Propre de Recherche 7061
}, CNRS Luminy, Case 907\\
F-13288 Marseille -- Cedex 9}

\vspace{3 cm}

{\huge\bf Semileptonic $B\to\rho$ and $B\to\pi$ Decays: Lattice and Dispersive
Constraints~\footnote{Talk
given at the High Energy Conference on Quantum
Chromodynamics (QCD 96), Montpellier, France, 4-12$^{th}$ July 1996. To appear
in the proceedings.}}

\vspace{0.7 cm}

{\Large\bf 
Laurent Lellouch}~\footnote{{\it email: lellouch@cpt.univ-mrs.fr}}

\vspace{2,3 cm}

{\bf Abstract}

\end{center}

I present two recent pieces of work on semileptonic $B\to\pi,\,\rho$
decays where it is shown how lattice QCD calculations can be used to test
heavy-quark symmetry and determine phenomenologically relevant
quantities despite the 
limits on these calculations' kinematical reach.  The
study of semileptonic $B\to\rho$ decays was performed with the UKQCD
Collaboration.

\vfill


\bigskip

\setcounter{footnote}{0}
\renewcommand{\thefootnote}{\arabic{footnote}}

\noindent Number of figures: 4

\bigskip

\noindent August 1996

\noindent CPT-96/P.3384
\bigskip

\noindent anonymous ftp or gopher: cpt.univ-mrs.fr
\end{titlepage}

\begin{abstract}
I present two recent pieces of work on semileptonic $B\to\pi,\,\rho$
decays where it is shown how lattice QCD calculations can be used to test
heavy-quark symmetry and determine phenomenologically relevant
quantities despite the 
limits on these calculations' kinematical reach.  The
study of semileptonic $B\to\rho$ decays was performed with the UKQCD
Collaboration.
\end{abstract}

\maketitle

\section{Motivations}

The CLEO Collaboration has very recently presented measurements of the
branching ratios for $B\to\rho\ell\bar\nu$ and $B\to\pi\ell\bar\nu$
decays ($\ell{=}e,\mu$)~\cite{cleo} and has announced its intention to
measure the corresponding differential decay rates.  These various
measurements represent an excellent opportunity to determine the
poorly known CKM matrix element $|V_{ub}|$. Such determinations require
understanding the non-perturbative, strong-interaction corrections to
the elementary $b-u-W$ coupling contained in the matrix elements of
the weak currents $V^\mu{=}\bar u\gamma^\mu b$ and $A^\mu{=}\bar
u\gamma^\mu\gamma^5 b$ between $B$ and $\pi$ or $\rho$ meson states.
It is to calculate these matrix elements that we resort to the
lattice.

$heavy\to light$ quark decays, such as the ones that concern us here,
are also interesting because they enable one to test heavy-quark
symmetry (HQS). For these decays, HQS is weaker than for $heavy\to
heavy$ quark decays: it only applies in a limited region around the
zero recoil point $q^2{=}q^2_{max}$, where $q$ is the four-momentum
transferred to the leptons, and imposes no normalization condition on
the relevant form factors at $q^2_{max}$. Nevertheless, because both
the mass and the spin of the heavy quark can be varied in lattice
calculations, the deviations from the heavy-quark limit due to finite
heavy-quark mass and spin effects can be measured.

\section{Limitations}

Current day lattice calculations, with lattice spacings on the order
of $3\gev^{-1}$, do not permit one to simulate $b\to light$ quark
decays over their full kinematical range. The problem is that the
energies and momenta of the particles involved, whose orders of
magnitude are set by the $b$ quark mass ($m_b\simeq 5\gev$), are large
on the scale of the cutoff in much of phase space.  
To limit these energies in relativsitc
lattice quark calculations, one performs the simulation with
heavy-quark mass values $m_Q$ around that of the charm ($m_c\simeq
1.5\gev$), where discretization errors remain under control. Then one
extrapolates the results up to $m_b$ by fitting heavy-quark scaling
relations (HQSR) with power corrections to the lattice results
(see \sec{hqex}).  Another approach is to work with discretized
versions of effective theories such as Non-Relativistic QCD (NRQCD) or
Heavy-Quark Effective Theory (HQET) in which the mass of the heavy
quark is factored out of the dynamics. All approaches, however,
are constrained to relatively small momentum transfers because of the
limited applicability of HQS and because of momentum-dependent
discretization errors.  So one can only reconstruct the $q^2$
dependence of the relevant form factors in a limited region around
$q^2_{max}$ and one is left with the problem of extrapolating these
results to smaller $q^2$.

$heavy\to light$ quark decays are difficult in any theoretical
approach. Indeed, they require understanding the underlying QCD dynamics
over a large range of momentum transfers from $q^2_{max}{=}26.4\gev^2
(20.3\gev^2)$ for semileptonic $B\to\pi$ ($B\to\rho$) decays, where
the final state hadron is at rest in the frame of the $B$ meson, to
$q^2{=}0$ where it recoils very strongly.

\section{$\bar B^0\to\rho^+\ell^-\bar\nu$ and a 
Model-Independent Determination of $|V_{ub}|$
\protect\footnote[1]
{The results presented here were obtained on a $24^3\times 48$ lattice 
at $\beta=6.2$ from 60 quenched 
configurations, using an $\ord{a}$-improved SW action for the quarks.
(Please see \protect\cite{JoF96} for details.)}
}
\label{sec:btorho}

One solution to the problem of the limited kinematical range of the
lattice results is to ignore the problem or rather rely on the
ingenuity of experimental groups to provide measurements of partial
rates in the region where lattice results are available. Combined with
lattice results for $B\to\rho\ell\bar\nu$ decays, such experimental
measurements will enable a model-independent determination of
$|V_{ub}|$~\cite{btorho}.  Rates should be sufficient since our
lattice results span a range of $q^2$ from $\sim 14.4\gev^2$ to
$q^2_{max}$ over which the partially integrated is
$4.6\er{4}{3}|V_{ub}|^2 ps^{-1}$. This represents approximatively
$1/3$ of the total rate obtained from light-cone sumrules (LCSR) in
\re{PBa96}, whose results at large $q^2$ agree well with ours.

\subsection{Form Factors and Heavy-Quark Extrapolation}
\label{hqex}

To describe $\bar B^0\to\rho^+\ell^-\bar\nu$ decays, we must evaluate
the matrix element $\la \rho^+(p',\eta)|V^\mu-A^\mu|\bar B^0(p)\ra$,
traditionally decomposed in terms of four form factors $A_1$,
$A_2$, $A$ and $V$ which are functions of $q^2$, where $q{=}p-p'$. We
calculate this matrix element for four 
values of the heavy-quark mass around that
of the charm. Then, to obtain $A_1$ at the scale of the
$B$ meson we fit, to the lattice results, the HQSR
\bea
A_1(\omega,M)\a_s(M)^{2/\beta_o}\sqrt{M}&= c(\omega)\big(1+
\frac{d(\omega)}{M}\nonumber\\
&+\ord{\Lambda^2/M^2}\big)
,\label{eq:hqsr}
\eea
where $M$ is the mass of the decaying meson, $\beta_0{=}11-2n_f/3$ and
$\Lambda$ is an energy characteristic of the light degrees of freedom.
This scaling relation holds for $\omega{=}(M^2+m^2-q^2)/2Mm$ close to
1 and the fit parameters $c$ and $d$ are independent of $M$. Here
$m$ is the mass of the final state meson.  Once $c$ and $d$ are fixed,
it is trivial to obtain $A_1(M{=}m_B)$ at the corresponding value of
$\omega$. Furthermore, $d$ determines the size of corrections to the
heavy-quark limit. Repeating this procedure for many values of
$\omega$, one obtains the $q^2$ dependence of the desired form factor
around $q^2_{max}$. The resulting $A_1(q^2,m_B)$ is plotted together
with the LCSR result of \re{PBa96} and the lattice results of
APE~\cite{ape} and ELC~\cite{elc}. Agreement amongst these four
calculations is excellent as it for $A_2$ and $V$~\cite{PBa96,JoF96},
which are obtained in an entirely analogous way.

\begin{figure}[tb]
\setlength{\epsfxsize}{60mm}\epsfbox[93 290 460 550]{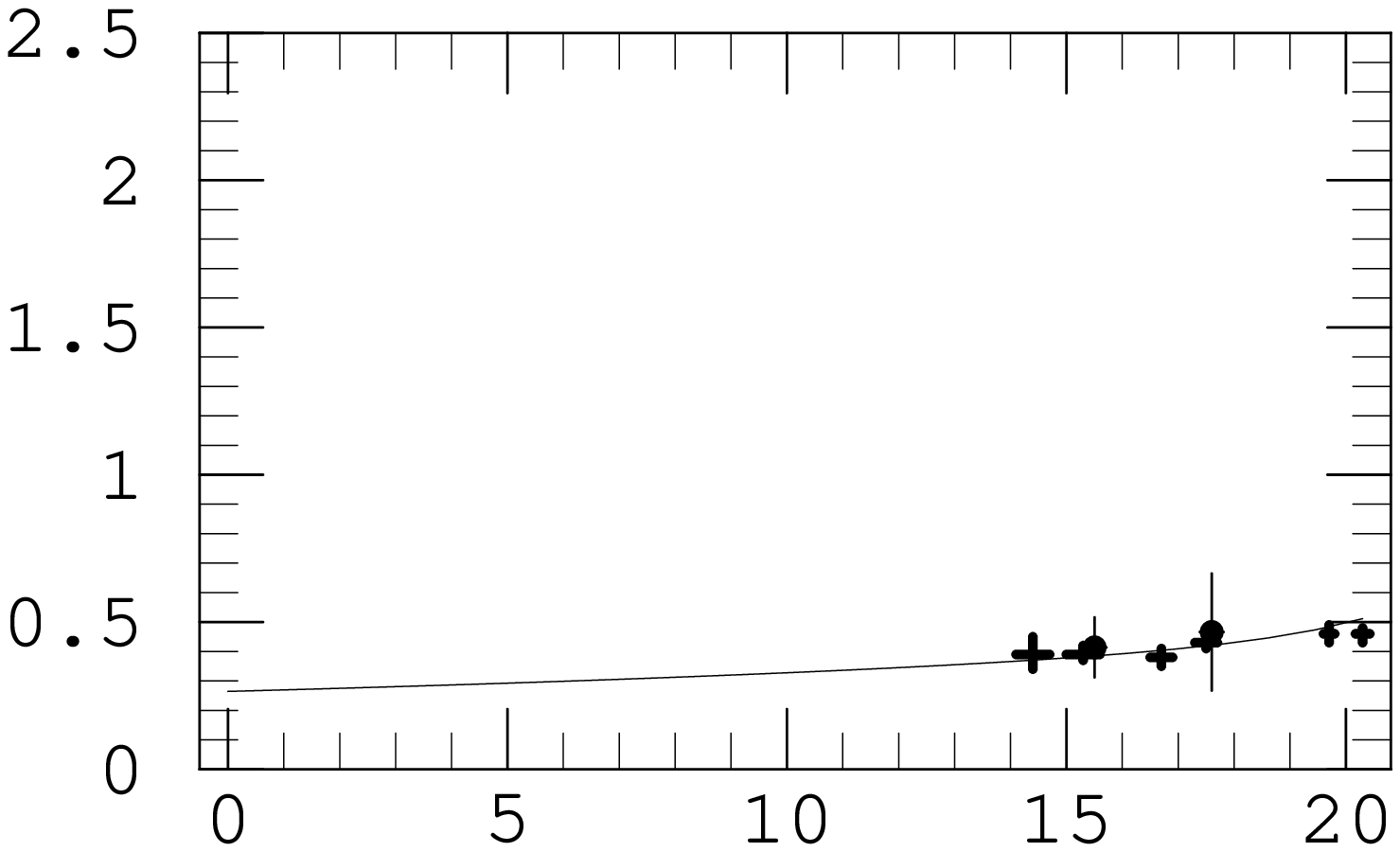}
\unit=0.8\hsize
\point 0.90 -0.00 {\large{$q^2(\mbox{GeV}^2)$}}
\point 0.30 0.55 {\Large{$A_1(q^2)$}}
\caption{$A_1$ vs. $q^2$ from UKQCD (crosses), 
APE\protect\cite{ape} (right-hand dot), ELC\protect\cite{elc}
(left-hand dot) and LCSR\protect\cite{PBa96} (curve). (Adapted from 
\protect \re{PBa96}.)}
\label{fig:comp}
\end{figure}

\subsection{Rates}

\begin{figure}[tb]
\setlength{\epsfxsize}{60mm}\epsfbox[30 100 500 530]{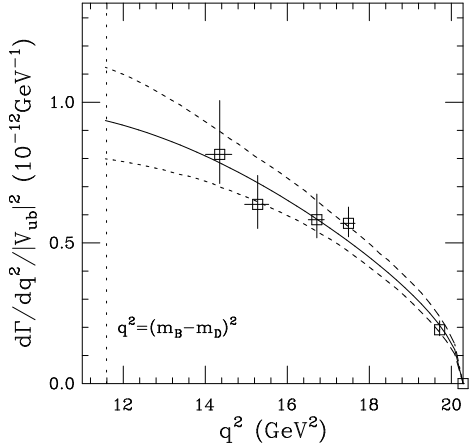}
\caption{The data points are our lattice results and the solid
curve, the fit to \protect\eq{eq:rate}.}
\label{fig:rate}
\end{figure}

Having determined $A_1$, $A_2$ and $V$, we can compute 
$(1/|V_{ub}|^2)$$d\Gamma/dq^2$. Our results are plotted in \fig{fig:rate}
(squares). In the region of $q^2$ accessed, we can
legitimately expand, around $q^2_{max}$,  the helicity amplitudes that
appear in the rate. Thus we fit, to the lattice points,
the parametrization
\bea
\frac{10^{12}}{|V_{ub}|^2}
\frac{d\Gamma}{dq^2}&\simeq&\frac{G^2_F}{192\pi^3m_B^3}
q^2\lambda(q^2)^{1/2} \nonumber\\
&\times& a^2\l(1+b(q^2-q^2_{max})\r)
\label{eq:rate}
\eea
where $\lambda(q^2){=}(M^2+m^2-q^2)-
4M^2m^2$. We find $a{=}4.6\er{4}{3}\pm
0.6 \gev$ and $b{=}(-8\er{4}{6}) 10^{-2} \gev^{-2}$ where the second
error on $a$ is systematic, all other errors being statistical.  With
$a$ and $b$ determined, the only unknown in \eq{eq:rate} is
$|V_{ub}|$. Therefore, a fit of the parametrization of \eq{eq:rate} to
an experimental measurement of the differential decay rate around
$q^2_{max}$ determines $|V_{ub}|$.  In this determination, $a$ plays
the role of ${\cal F}(1)$ in the extraction of $|V_{cb}|$ from
semileptonic $B\to D$ or $D^*$ decays \cite{MNe94} and $b$ the role
of ${\cal F}'(1)$. The difference,
here, is that $a$ is not determined by HQS up to small radiative and
power corrections. It is a genuinely non-perturbative quantity. 
Another way of determining $|V_{ub}|$ from the lattice results is to
compare partially integrated rates from $q^2\ge 14\gev^2$ to $q^2_{max}$
given by \eq{eq:rate} to the corresponding experimental measurements.
Both these methods yield $|V_{ub}|$ with approximatively 10\% statistical
and 12\% theoretical uncertainties.

\subsection{A Test of HQS}

\begin{figure}[tb]
\setlength{\epsfxsize}{30mm}\epsfbox[170 290 285 690]{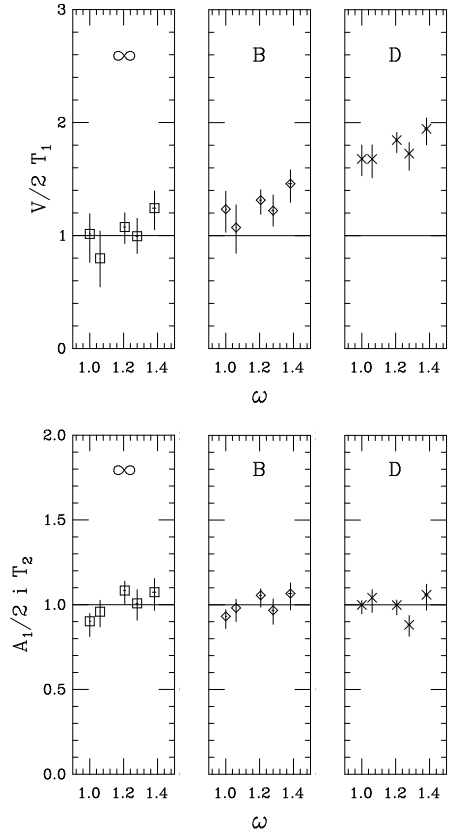}
\caption{Ratios $V/2T_1$ and $A_1/2iT_2$ for 5 values of $\omega$ and three
initial meson masses. The solid lines are the HQS predictions.}
\label{fig:sem_over_rad}
\end{figure}

In \fig{fig:sem_over_rad} we compare semileptonic $B\to\rho$ form factors
with those governing the short distance contribution to radiative
$B\to K^*\gamma$ decays for which the relevant hadronic matrix element
is $\la K^*(p',\eta)|\bar
s\sigma^{\mu\nu}q^\nu b_R|B(p)\ra$, with $q{=}p-p'$. This
matrix element is traditionally decomposed in terms of three form
factors, $T_1$, $T_2$ and $T_3$. The comparison is made for three
initial meson masses: $M{=}m_D$, $M{=}m_B$ and $M\to\infty$.
For identical final-state vector
mesons (in
\fig{fig:sem_over_rad} all light-quarks involved have the same mass,
slightly larger than that of the strange), HQS predicts
\be
V(q^2)=2T_1(q^2),\quad A_1=2iT_2(q^2)
\ ,\label{eq:sor}
\ee
for $q^2$ around $q^2_{max}$ or, equivalently,
$\omega$ close to 1. 
While $V/2T_1$ displays large
$1/M$ corrections at the $D$ and even $B$ meson scale, $A_1/2iT_2$
exhibits no such corrections even at the $D$ scale. 
Both ratios, however, converge to 1 in the heavy-quark limit
which gives us confidence that we control the heavy-quark-mass dependence
of the various form factors. Furthermore, 
these ratios can help constrain the possible $q^2$ dependences
of the various form factors around $q^2_{max}$ at $M{=}m_B$.


\section{$\bar B^0\to\pi^+\ell^-\bar\nu$ and Dispersive Constraints}

A second solution to the problem of the limited kinematical reach of
lattice simulations of $heavy\to light$ quark decays is
to combine lattice results for the relevant form factors around
$q^2_{max}$ with dispersive bound techniques to obtain improved,
model-independent bounds for the form factors for all $q^2$~\cite{btopi}. 
For the
case of $\bar B^0\to\pi^+\ell^-\bar\nu$ decays, whose hadronic matrix
element, $\la \pi^+(p')|V^\mu|\bar B^0(p)\ra$, is traditionally decomposed 
in terms of two form factors $f^+(q^2)$ and $f^0(q^2)$, one can use the
kinematical constraint, $f^+(0)${=}$f^0(0)$, to further constrain the
bounds.

\subsection{Dispersive Bounds}

The subject of dispersive bounds in semileptonic decays 
has a long history going back to S. Okubo {\it et al.}~who
applied them to semileptonic $K\to\pi$ decays~\cite{SOkS71}. 
C. Bourrely {\it et al.}~first combined these techniques with
QCD and applied them to semileptonic $D\to K$ decays \cite{CBoMR81}. 
Very recently,
C.G. Boyd {\it et al.}~applied them to $B\to\pi\ell\bar\nu$
decays \cite{CBoGL95}. 

The starting point for $B\to\pi\ell\bar\nu$ decays is the polarization
function
\bea
\Pi^{\mu\nu}(q){=}i\int d^4x\ e^{iq\cdot x} \la 0|T\l(V^\mu(x)
V^{\nu\dagger}(0)\r)|0\ra\nonumber\\
{=}(q^\mu q^\nu-g^{\mu\nu}q^2)\,\Pi_T(q^2)+q^\mu q^\nu \,\Pi_L(q^2)
\ ,\label{twopoint}
\eea
where $\Pi_{T(L)}$ corresponds to the propagation of a
$J^P{=}1^-\,(0^+)$ particle. The corresponding spectral functions,
$\im\Pi_{T,L}$, are sums of positive contributions coming from
intermediate $B^*$ ($J^P{=}1^-$), $B\pi$ ($J^P{=}0^+\mbox{ and }1^-$),
$\ldots$ states and are thus upper
bounds on the $B\pi$ contributions. 
Combining, for instance, the bound from $\im\Pi_L$ with the
dispersion relation ($Q^2{=}-q^2$)
\bea
\chi_L(Q^2)&=&\frac{\partial}{\partial Q^2} (Q^2\Pi_L(Q^2))\nonumber\\ 
&=&\frac{1}{\pi}\int_0^\infty
dt\frac{t\,\im\Pi_L(t)}{\l(t+Q^2\r)^2}
\ ,\label{eq:disprelS}
\eea
one finds
\bea
\chi_L(Q^2)\ge\frac{1}{\pi}\int_{t^+}^\infty
dt\,k(t,Q^2)|f^0(t)|^2
\ ,\label{eq:chilbnd}
\eea
where $t_{\pm}{=}(m_B\pm m_\pi)^2$ and $k(t,Q^2)$ is a kinematical
factor.  Now, since $\chi_L(Q^2)$ can be calculated analytically in
QCD for $Q^2$ far enough below the resonance region (i.e. $-Q^2\ll
m_b^2$), \eq{eq:chilbnd} gives an upper bound on the weighted integral
of the magnitude squared of the form factor $f^0$ along the $B\pi$
cut. To translate this bound into a bound on $f^0$ in the region of
physical $B\to\pi\ell\bar\nu$ decays is a problem in complex
analysis (please see \re{btopi} for details).
A similar constraint can be obtained from $\Pi_T$ for $f^+$. There, however,
one has to confront the additional difficulty that $f^+$ is not analytic
below the $B\pi$ threshold because of the $B^*$ pole.

The beauty of the methods of \re{CBoMR81} is that they enable one to
incorporate information about the form factors, such as their
values at various kinematical points, to constrain the bounds.
For the case at hand, however, these methods must be generalized in two
non-trivial ways. In constructing these generalizations, one must
keep in mind that the bounds: 1) form inseparable pairs; 2) do not 
indicate the probability that the form factor
will take on any particular value within them.

\subsection{Imposing the Kinematical Constraint}

The first problem is that \eq{eq:chilbnd} and the equivalent
constraint for $f^+$ yield independent bounds on the form
factors which do not satisfy the kinematical constraint 
$f^+(0){=}f^0(0)$. The bounds on $f^+$ require $f^+(0)$ to lie within
an interval of values $I_+$ and those on $f^0$, within an interval
$I_0$.  Together with these bounds, however, the kinematical constraint 
requires $f^+(0){=}f^0(0)$ to lie somewhere within $I_+\cap I_0$.
Thus, we seek bounds on the form factors which are consistent
with this new constraint.

A natural definition is to require these new bounds to be the envelope
of the set of pairs of bounds obtained by allowing $f^+(0)$ and
$f^0(0)$ to take all possible values within the interval $I_+\cap
I_0$.  In \re{btopi}, I show how this envelope can be constructed
efficiently and that the additional constraint can only improve the
bounds on the form factors for all $q^2$. Also, as a by product, one
obtains a formalism which enables one to constrain bounds on a
form factor with the knowledge that it must lie within an interval of
values at one or more values of $q^2$.

\subsection{Taking Errors into Account}

As they stand, the methods of \re{CBoMR81} can only accommodate exact
values of the form factors at given kinematical points and contain no
provisions for taking errors on these values into account. Of course,
the results given by the lattice do carry error bars. More precisely,
the lattice provides a probability distribution for the value of the
form factors at various kinematical points. What must be done, then,
is to translate this distribution into some sort of probability
statement on the bounds.
The conservative solution is to consider the probability that complete
pairs of bounds lie within a given finite interval at each value of
$q^2$. Then, using this new probability, one can define upper and
lower $p\%$ bounds at each $q^2$ as the upper and lower boundaries of
the interval that contains the central $p\%$ of this 
probability.\footnote{The density of pairs of bounds increases 
toward the center 
of the ``distribution'' as long as the distribution of the lattice results
does.} These bounds indicate that there is at least a $p\%$ probability that
the form factors lie within them at each $q^2$.

\subsection{Lattice-Constrained Bounds}

\begin{figure}[tb]
\setlength{\epsfxsize}{64mm}\epsfbox[80 300 470 545]{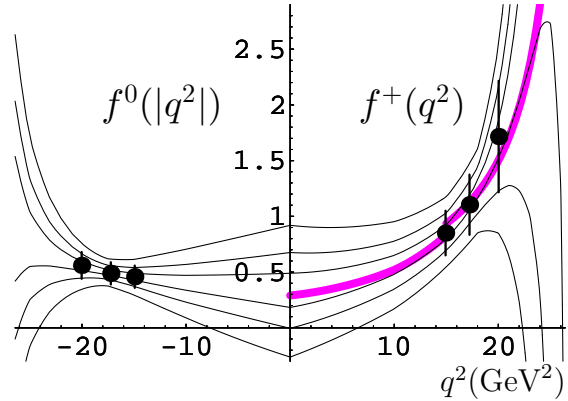}
\unit=0.8\hsize
\point 0.95 -0.07 {\large{$q^2(\mbox{GeV}^2)$}}
\point 0.20 0.55 {\Large{$f^0(|q^2|)$}}
\point 0.770 0.55 {\Large{$f^+(q^2)$}}
\caption{$f^0(|q^2|)$ and $f^+(q^2)$ versus $q^2$.
The data points are the lattice results of UKQCD\protect\cite{DBuetal95} 
with added systematic errors. 
The pairs of fine curves are, from the outermost
to the innermost, the 95\%, 70\% and 30\% bounds. The shaded curve is 
the LCSR result of \protect\re{VBeBKR95}.}
\label{fig:lcsr}
\end{figure}

To constrain the bounds on $f^+$ and $f^0$, I use the lattice
results of the UKQCD Collaboration
\cite{DBuetal95}, to which I add a large range of systematic
errors to ensure that the bounds obtained are conservative. Because of
these systematic errors, the probability distribution of the lattice
results is not known. I make the simplifying and rather conservative
assumption that the results are uncorrelated and gaussian distributed.
I construct the required probability by generating 4000 pairs of
bounds from a Monte-Carlo on the distribution of the lattice results.
My results for the bounds on the form factors are shown in
\fig{fig:lcsr}. I have plotted the two form factors back-to-back to
show the effect of the kinematical constraint. Without it, the bounds
on $f^+$ would be looser, especially around $q^2{=}0$, where phase
space is large. Since $f^+$ determines the rate, the kinematical 
constraint and the bounds on $f^0$ are important.

Also shown in \fig{fig:lcsr} is the LCSR result of \re{VBeBKR95} which
has two components: for $q^2$ below $15\gev^2$, the $q^2$ dependence
of $f^+$ is determined directly from the sumrule; for larger $q^2$,
pole dominance is assumed with a residue determined from the same
correlator. Agreement with the bounds is excellent. In \re{btopi}, the
bounds are compared with the predictions of more authors as well as
with direct fits of various parametrizations to the lattice results.
Though certain predictions are strongly disfavored, the lattice results
and bounds will
have to improve before a firm conclusion can be drawn as to the
precise $q^2$ dependence of the form factors.

The bounds on $f^+$ also enable one to constrain the $B^*B\pi$
coupling $g_{B^*B\pi}$ which determines the residue of the $B^*$ pole
contribution to $f^+$.
The constraints obtained are poor because $f^+$ is weakly bounded at
large $q^2$. Fitting the lattice results for $f^0$ and $f^+$ to a
parametrization which assumes $B^*$ pole dominance for $f^+$ and which
is consistent with HQS and the kinematical constraint gives the more
precise result $g_{B^{*+}B^o\pi^+}=28\pm 4$.\footnote{The result of 
this fit is entirely compatible with our bounds
on $f^+$ and $f^0$.}
However, because this
result is model-dependent, it should be taken with care.

\subsection{Bounds on the rate and $|V_{ub}|$}

As was done for the form factors, one can define the probability of
finding a complete pair of bounds on the rate in a given interval and
from that probability determine confidence level (CL) intervals for the
rate.  The resulting bounds are summarized in \tab{tab:rate}. They
were obtained by appropriately integrating the 
4000 bounds generated for $f^+(q^2)$, 
taking the skewness of the resulting ``distribution'' 
of bounds on the rate into account. 
\begin{table}[tb]
\caption{Bounds on rate in units of $|V_{ub}|^2\,ps^{-1}$ and on $f^+(0)$.
\label{tab:rate}}
\begin{tabular}{ccc}
\hline
$\Gamma\l(\bar B^0\to\pi^+\ell^-\bar\nu\r)$ & $f^+(0)$ & CL\\
\hline
$2.4\to 28$ & $-0.26\to 0.92$ & 95\% \\
$2.8\to 24$ & $-0.18\to 0.85$ & 90\% \\
$3.6\to 17$ & $0.00\to 0.68$ & 70\% \\
$4.4\to 13$ & $0.10\to 0.57$ & 50\% \\
$4.8\to 10$ & $0.18\to 0.49$ & 30\% \\
\hline
\end{tabular}
\end{table}
The CL 
bounds obtained can be used, in conjunction with the branching ratio 
measurement of CLEO \cite{cleo}, to determine $|V_{ub}|$. One finds
\be
|V_{ub}|10^4\sqrt{\tau_{B^0}/1.56\,ps}=(34\div 49)\pm 8\pm 6
\ ,\label{eq:vub}
\ee
where the range given in parentheses is that obtained from the
30\% CL bounds on the rate and represents the most probable
range of values for $|V_{ub}|$. The first set of errors is obtained
from the 70\% CL bounds and the second is obtained by
combining all experimental uncertainties in quadrature and applying them to
the average value of $|V_{ub}|$ given by the 30\% CL results.
This determination of $|V_{ub}|$ has a theoretical error of approximately
37\%. Though non-negligible, this error is quite reasonable given
that the bounds on the rate are completely model-independent
and are obtained from lattice data which lie in a limited kinematical domain 
and include a conservative range of systematic errors.

\section{Conclusion and Outlook}

Because HQS applies to $heavy\to light$ quark decays in a rather
limited way, it is not possible to determine the full $q^2$ dependence
of the relevant form factors from the lattice alone. The flip side of
the coin is that the model-independent information provided by lattice
calculations about these decays, though limited, is still very
important, because the relevant matrix elements are not anchored at zero
recoil by HQS, up to small radiative and power corrections, as they
are in $heavy\to heavy$ quark decays. 

I have presented two approaches by which the information provided
by the lattice on exclusive semileptonic $b\to u$ decays can
be used to extract $|V_{ub}|$. Both approaches 
will benefit from forthcoming, improved
lattice results. The lattice-constrained bounds would also benefit 
enormously from an increase in the range of accessible $q^2$. 

Finally, the techniques developed in \re{btopi} to construct lattice-improved
bounds for $B\to\pi\ell\bar\nu$ decays 
are in principle applicable to limited results obtained by
non-lattice means and to other processes such as $B\to\rho\ell\bar\nu$
and $B\to K^*\gamma$ decays.

\end{document}